\documentclass[reprint,amsmath,amssymb,aps,prl]{revtex4-1}
\usepackage[dvips]{graphicx}
\usepackage{dcolumn}
\usepackage{bm}
\usepackage{hyperref}
\usepackage{color}
\usepackage{cancel}
\usepackage{bmpsize}
\usepackage{amsmath}
\usepackage{amssymb}

\newcommand{\be}[0]{\begin{equation}}
\newcommand{\ee}[0]{\end{equation}}

\newcommand{\lra}\simeq

\definecolor{linkcolor}{rgb}{0.9,0,0}
\definecolor{citecolor}{rgb}{0,0.6,0}
\definecolor{urlcolor}{rgb}{0,0,1}
\hypersetup{
	colorlinks, linkcolor={linkcolor},
	citecolor={citecolor}, urlcolor={urlcolor}
}

\begin{document}

\title{Synthesis of the Einstein-Podolsky-Rosen entanglement \\in a sequence of two single-mode squeezers}

\author{Ilya A. Fedorov$^{1,2}$, Alexander E. Ulanov$^{1,3}$, Yury V. Kurochkin$^1$ and A.I. Lvovsky$^{1,2,4}$}

\affiliation{$^1$Russian Quantum Center, 100 Novaya St., Skolkovo, Moscow 143025, Russia}
\affiliation{$^2$P. N. Lebedev Physics Institute, Leninskiy prospect 53, Moscow 119991, Russia}
\affiliation{$^3$Moscow Institute of Physics and Technology, 141700 Dolgoprudny, Russia}
\affiliation{$^4$Institute for Quantum Science and Technology, University of Calgary, Calgary AB T2N 1N4, Canada}

\email{trekut@gmail.com}
\date{\today}
%
\begin{abstract}

Synthesis of the Einstein-Podolsky-Rosen entangled state --- the primary entangled resource in  continuous-variable quantum-optical information processing --- is a technological challenge of great importance. Here we propose and implement a new scheme of generating this state. Two nonlinear optical crystals, positioned back-to-back in the waist of a pump beam, function as single-pass degenerate optical parametric amplifiers and produce single-mode squeezed vacuum states in orthogonal polarization modes, but in the same spatiotemporal mode. A subsequent pair of waveplates acts as a beam splitter, entangling the two polarization modes to generate the Einstein-Podolsky-Rosen state. This technique takes advantage of the strong nonlinearity associated with type-I phase-matching configuration while at the same time eliminating the need for actively stabilizing the optical phase between the two squeezers, which typically arises if these squeezers are spatially separated. We demonstrate our method in an experiment, preparing a 1.4 dB two-mode squeezed state and characterizing it via two-mode homodyne tomography. 

\end{abstract}

\maketitle
\vspace{10 mm}

\textit{Introduction.} The two-mode squeezed vacuum state is characterized by simultaneous correlation of the positions and anticorrelation of momenta of two harmonic oscillators beyond the standard quantum limit defined by the vacuum state. Discovered in 1935 by Einstein, Podolsky and Rosen \cite{EPR1935}, this state (which we hereafter refer to as the EPR state) gave rise to the celebrated quantum nonlocality paradox. With the emergence of quantum-optical information technology, the EPR state became the primary entangled resource for the continuous-variable domain. Its applications include quantum information processing \cite{BRA05,CVreview1}, quantum metrology \cite{EPR_metrology}, quantum cryptography \cite{Ralph1999,EPR_QKD}, teleportation \cite{Furusawa1998} and quantum repeater \cite{Kurochkin_distillation, Ulanov_distillation} protocols.

While the EPR state can be prepared in a variety of nonlinear physical media, such as atomic ensembles \cite{Marino2008} and fibers \cite{Silberhorn2001} , the most common approach employs spontaneous parametric down-conversion (SPDC) in a crystal with second-order nonlinearity. If SPDC is utilized in the non-degenerate configuration, the EPR state is produced directly in its two output channels \cite{Ou92,Zhou2015}. This approach, however, suffers from a few shortcomings. In particular, if the SPDC is non-collinear, critical phase matching may complicate the optical alignment or distort the emission modes. Furthermore, such an arrangement is difficult to set up within a Fabry-Perot resonator. On the other hand, settings in which the emission modes  are instead separated by polarization by means of the type-II SPDC configuration are of disadvantage with respect to type-I because of significantly lower optical nonlinearity \cite{Dmitriev1993}. 

A frequently employed alternative approach is to first obtain single-mode position- and momentum-squeezed states in two independent optical modes via degenerate SPDC, and subsequently make them interfere on a symmetric beam splitter \cite{Furusawa1998,Takeda2013,Eberle2013}. The quadrature operators of the two modes then transform according to 
\begin{equation}
\label{eq2}
\left[ {\begin{array}{c}
	X_1' \\
	X_2' \\
	\end{array} } \right]
=
\dfrac{1}{\sqrt{2}}
\left[ {\begin{array}{cc}
	1 & -1 \\
	1 & 1 \\
	\end{array} } \right]
\left[ {\begin{array}{c}
	X_1 \\
	X_2 \\
	\end{array} } \right],
\end{equation}
producing the two-mode squeezed state at the output. While this method allows one to obtain high degrees of squeezing \cite{Eberle2013}, it requires precise phase locking between the two squeezed light sources, which is a significant technical complication.

\begin{figure}[b]
	\includegraphics[width=3.5in]{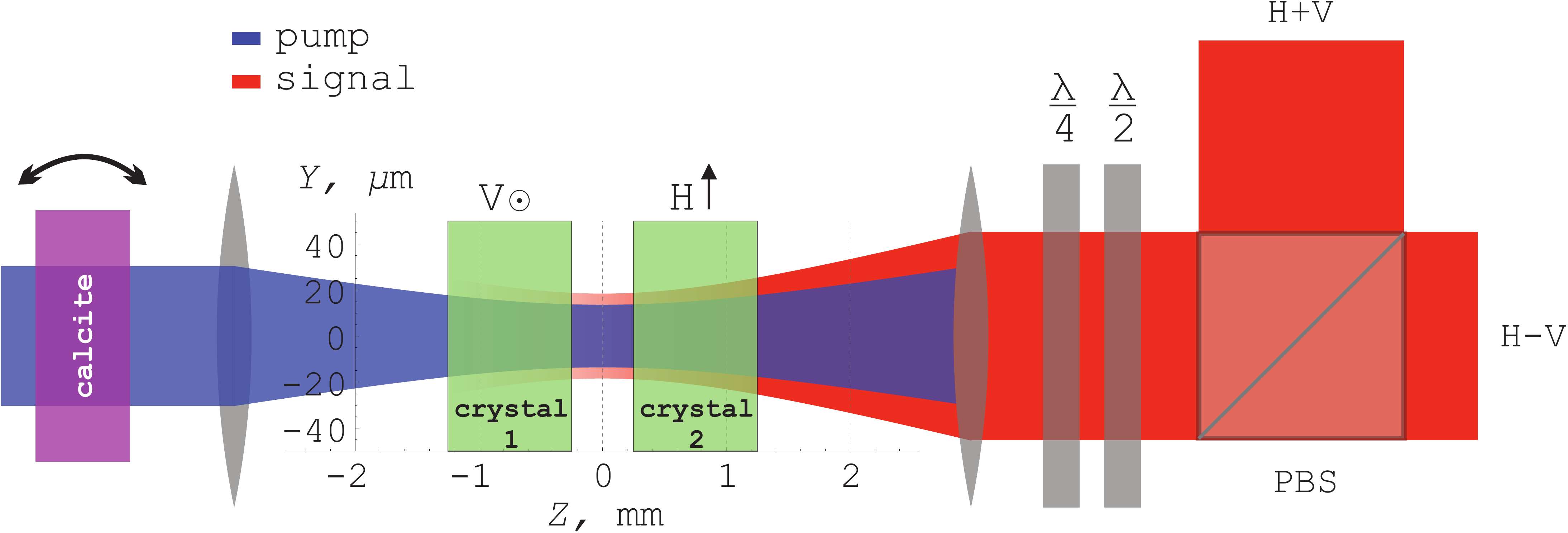}
	\caption{Schematic of the setup. Both crystals are aligned for collinear, type-I phase matching conditions. PBS, polarizing beam splitter. The axes illustrate the longitudinal beam waist behavior and the positioning of the crystals with respect to that waist. Other elements are not drawn to scale.}
	\label{p1}
\end{figure}

In this work, we propose a scheme devoid of these drawbacks. The idea, inspired by Kwiat {\it et al.} \cite{Kwiat1999}, is to use a series of two degenerate SPDC processes, placing the two nonlinear crystals in a single waist of the pump beam immediately one after another (Fig.~\ref{p1}). If the optical axes of the crystals are orthogonal to each other, and the pump beam is polarized diagonally between them, the two nonlinear processes will squeeze orthogonally polarized vacuum modes. The two squeezed vacua thereby populate a single spatial mode, and can be put to interference using a pair of waveplates. Since the distance between the crystals is limited by a few millimeters, air density fluctuations between the two squeezed vacua are negligible, so no phase locking is required.

\textit{Experiment.}
Both down-conversion processes take place in periodically poled potassium titanyl-phosphate crystals. The first and second crystals are phase matched for type I degenerate SPDC into the vertical and horizontal polarization modes, respectively. The  phase matching is aligned by independent angle and temperature control of each crystal. 
The crystals are pumped in a single-pass manner with frequency-doubled pulses at $\lambda = 390$ nm, generated by a Ti:Sapphire laser with a repetition rate of 76 MHz and a pulse width of 1.5 ps \cite{instantFock}. The average power of the pump field is 80 mW. The polarization of the pump field is diagonal, so the pump intensity is distributed equally to both nonlinear processes.

\begin{figure}[t]
\includegraphics[width=3.4in]{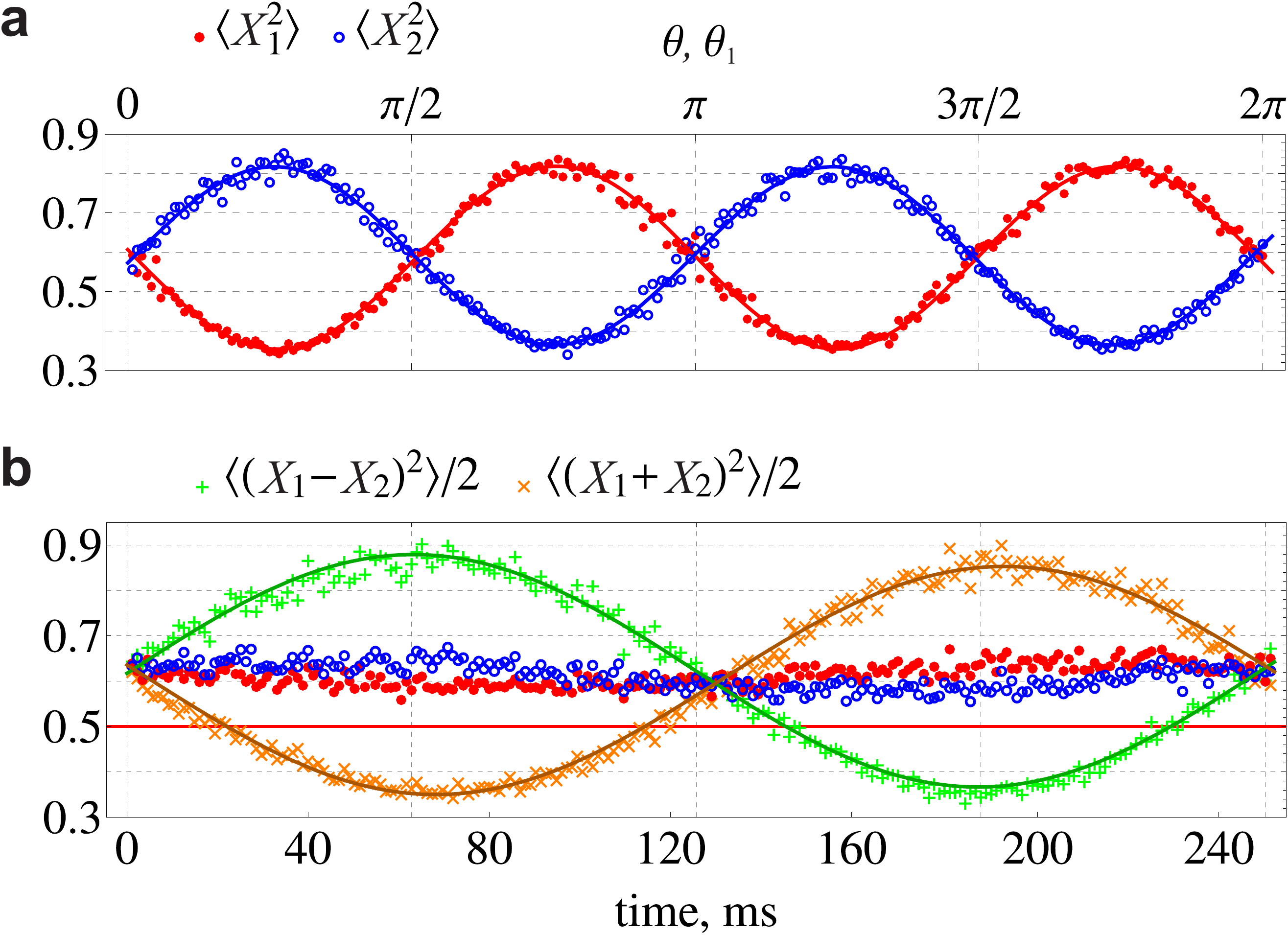}
\caption{
\textbf{a.} Variances of the initial, orthogonally-squeezed single-mode vacua as a function of time for a linearly varying quadrature phase $\theta$, before the interference. Horizontally (vertically) polarized modes: open (filled) circles. Lines show the theoretical prediction (\ref{single}) with $\zeta = 0.44$, detected with efficiency 52\%. Data are normalized so that the vacuum noise level is 0.5. 
\textbf{b.} Variances of the EPR-state quadratures as a function of time. In this experiment, the phase of one of the LOs is varied, $\theta_1$ in (\ref{eqEPR}). Circles: individual quadratures. Diagonal/vertical crosses: sum and difference quadratures.
Lines show the theoretical prediction for the EPR state (\ref{eqEPR}) with $\zeta = 0.44$ detected with a 50\% efficiency. The minimum difference-quadrature variance is 0.36, which corresponds to 1.4 dB of two-mode squeezing.}
\label{p2}
\end{figure}

The length of the crystals along the beam is 1 mm, as shown in Fig.~\ref{p1}. Both crystals are set in the waist of the focused pump beam with the Gaussian radius of $w_0 = 12.4 \,\mu$m, with corresponding Rayleigh range being $z_R = \pi w_0^2/\lambda = 1.25$ mm. The distance between the crystals is $0.45$ mm, so that their centers are $0.72$ mm away from the beam waist; the beam width at the center points is then $1.15 w_0$.

After the second crystal, two orthogonally-polarized squeezed vacua, sharing the same spatial mode, are prepared. To characterize these states, the modes are separated by a polarizing beamsplitter (Fig.~\ref{p1}) and subjected to homodyne measurement \cite{Kumar2012}. 
The variance of the quadrature data from both squeezed states as a function of phase and time is shown in the Fig.~\ref{p2}(a). The sinusoidal behavior of the variance is due to linear variation of the optical phase $\theta$ of the local oscillator \cite{Lvovsky2009}. This behavior is modeled by a perfect squeezer with squeezing parameter $\zeta$ followed by a loss channel with transmissivity $\eta$, in which the quadrature variance is given by \cite{LvovskySQ}
\begin{equation}\label{single}
\langle X_\theta^2\rangle=\frac\eta 2(\cosh 2\zeta-\cos 2\theta\sinh 2\zeta)+\frac{1-\eta}2.
\end{equation}
The single-mode experimental data can be fit with this equation using  $\zeta = 0.44$, $\eta = 0.52$ [Fig.~\ref{p2}(a)].

The single-mode squeezed states generated in the two crystals are delayed with respect to one another due to the difference in group velocities of the pump and the signal. The experimentally observed delay of $0.58$ mm is in agreement with the prediction  based on the Sellmeyer model: \cite{Bierlein1989}
\begin{equation}
\label{eq4}
v_{\lambda} = 0.41 c, \quad v_{2\lambda} = 0.52 c,
\end{equation}
with $c$ being the speed of light in vacuum. 
To compensate for that difference, we retard the vertical polarization component of the pump with respect to the horizontal one, using a birefringent calcite crystal of length 3.6 mm.
The birefringence of calcite depends on its orientation. We rotate the crystal around the vertical axis, as shown in Fig.~\ref{p2}(a), to optimize the delay between the polarization components of the pump and to set the phase between the two squeezed vacua to $\pi/2$.

The two squeezed states are subsequently brought to interference. The latter is realized in the polarization basis by means of a  half-wave plate with its optical axis at 22.5$^\circ$ to horizontal, and a quarter-wave plate for fine tuning. 
The interfered modes are spatially separated on the polarizing beamsplitter and directed to homodyne detectors \cite{Kumar2012}.

\begin{figure}[t]
\includegraphics[width=3.4in]{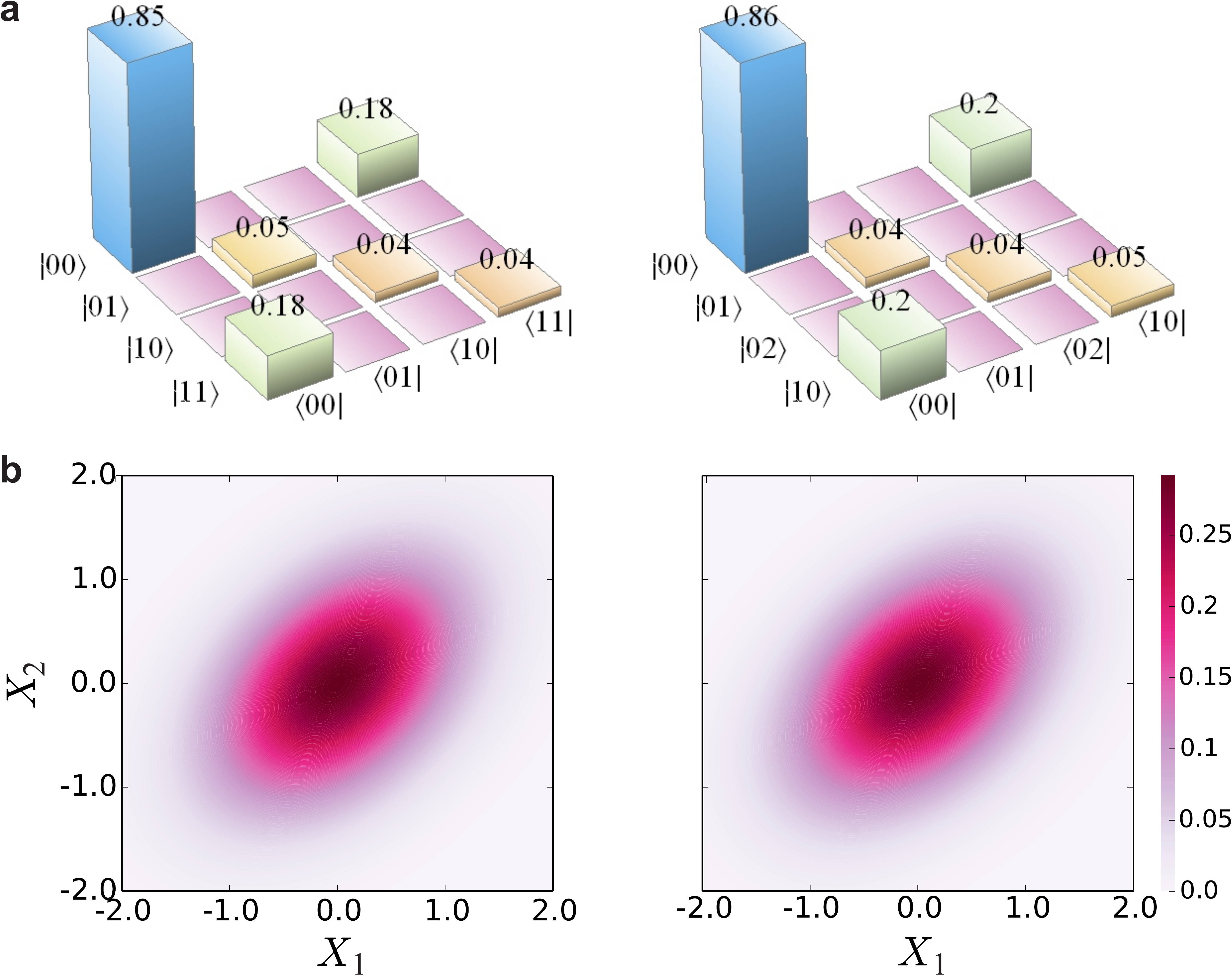}
\caption{Reconstructed EPR state (left) and theoretical expectation, based on the experimental input single-mode squeezed states, right. \textbf{a.} Density matrices in Fock basis. \textbf{b.} Correlated probability densities for the position quadratures.}
\label{p3}
\end{figure}

The quadrature variances from each individual mode are shown in Fig.~\ref{p2}(b) in blue and purple. Both exhibit approximately constant variance as expected for the EPR state, whose individual modes considered separately are in a thermal state \cite{LvovskySQ}. The sum- and difference- quadrature variances, shown in green and orange, demonstrate opposite phase-dependent variances characteristic of the EPR state. The theoretical expectation for these variances is \cite{Kurochkin2014}
\begin{eqnarray}
\left\langle \frac {[X_{1,\theta_1} \mp X_{2,\theta_2}]^2]}2\right \rangle=&
\label{eqEPR}
\\ &\hspace{-3cm}=  \frac\eta 2[\cosh(2\zeta)\pm \cos(\theta_1+\theta_2)\sinh(2\zeta)] +\frac{1-\eta}2 \nonumber
\end{eqnarray}
and the best fit is obtained with $\zeta = 0.44$, $\eta=0.50$. The small decrease in the efficiency with respect to the single-mode case is attributed to the non-ideal interference of the input states, which also explains the residual phase dependence of the individual quadrature variances, visible in the Fig.~\ref{p2}(b). Note that, compared to the single-mode squeezed vacua [Fig.~\ref{p2}(a)], the two-mode quadrature variances oscillate at half frequency, as expected from Eqs.~(\ref{single}) and (\ref{eqEPR}).

The result of two-mode homodyne tomography of the EPR state \cite{MaxLik07} is presented in the Fig.~\ref{p3}, left. The mean photon population of each mode is $0.11$, which is in agreement with the expected value $\eta\zeta^2=0.10$. The right side of Fig.~\ref{p3} shows the theoretical expectation \cite{LvovskySQ} with the squeezing and efficiency correction for $\eta=0.5$. Its fidelity with the experimentally reconstructed state is 98\%.

\textit{Summary.}
We demonstrated preparation of the two-mode squeezed vacuum state in a series of two type-I nonlinear crystals prositioned back-to-back in a waist of the pump beam. The demonstrated technique can be seen as continuous-variable analog of the method for generating polarization-entangled photon pairs \cite{Kwiat1999}. It takes advantage of the relative strength of optical nonlinearity in type-I SPDC while eliminating the need for critical phase-matching  and the phase locking of two independent squeezers.
It can be adapted to cavity- or fiber-based optical parametric amplifier schemes.


\vspace{1em}
\noindent
The support from the Ministry of Education and Science of the Russian Federation in the framework of the Federal Program (Agreement 14.582.21.0009) is gratefully acknowledged.

\end{document}